%% file: main.tex
\documentclass[aps,prd,preprint,tightenlines,superscriptaddress,showpacs,byrevtex]{revtex4}
\usepackage{graphicx} % Include figure files
\usepackage{dcolumn}  % Align table columns on decimal point
\usepackage{amsmath}
\usepackage{xspace}
\usepackage{SIunits}

\newcommand{\B}{\ensuremath{B}\xspace}
\newcommand{\Bz}{\ensuremath{\B^0}\xspace}
\newcommand{\BzB}{\ensuremath{\bar{\B}{}^0}\xspace}
\newcommand{\ups}{\ensuremath{\Upsilon(4S)}\xspace}
\newcommand{\jpsi}{\ensuremath{J\!/\psi}\xspace}

\newcommand{\dz}{\ensuremath{\Delta z}\xspace}
\newcommand{\dt}{\ensuremath{\Delta t}\xspace}
\newcommand{\CT}{\ensuremath{CT}\xspace}
\newcommand{\WT}{\ensuremath{WT}\xspace}

\newcommand{\taub}{\ensuremath{\tau_{\Bz}}\xspace}
\newcommand{\dm}{\ensuremath{\Delta m_d}\xspace}
\newcommand{\dg}{\ensuremath{\Delta \Gamma}\xspace}
\newcommand{\Asl}{\ensuremath{A_{sl}}\xspace}

\newcommand{\cm}{\centi\metre\xspace}
\newcommand{\micron}{\micro\metre\xspace}

\newcommand{\invfb}{\femto\reciprocal\barn\xspace}

\newcommand{\gevc}{\giga\electronvolt\!/c\xspace}
\newcommand{\gevcc}{\giga\electronvolt\!/c^2\xspace}
\newcommand{\mev}{\mega\electronvolt\xspace}

\newcommand{\mevcc}{\mega\electronvolt\!/c^2\xspace}

\begin{document}

\preprint{\vbox{ \hbox{   }
                 \hbox{BELLE-CONF-0451}
                 \hbox{ICHEP04 8-0699} 
}}

\title{ \quad\\[0.5cm]
$CP$ violation in $B$-$\bar{B}$ mixing with dilepton events}
\date{\today}
\include{author-conf2004}

\begin{abstract}
We report a measurement of the charge asymmetry for same-sign dileptons  
in \Bz-\BzB mixing. 
The data were collected with the Belle detector at KEKB. Using a data sample of 
78 \invfb recorded at the \ups resonance and 
9 \invfb recorded at an energy 60 \mev  below the resonance, we measure 
$\Asl = ( -0.13 \pm 0.60(\text{stat}) \pm 0.56(\text{sys}) )\%$. 
\end{abstract}


\maketitle

\section{Introduction}

The standard model allows $CP$-violation in 
\Bz-\BzB mixing~\cite{cpvmix}.  In particular, there is a possible difference 
between the $\Bz \to \BzB$ and  $\BzB \to \Bz$ transition
rates, which can manifest itself as a charge asymmetry in the same-sign dilepton events in 
\ups decays when prompt leptons from
semileptonic decays of neutral \B mesons are selected. 
With the assumption of \emph{CPT} invariance in the mixing, the flavor and
mass eigenstates of the neutral \B mesons are related by 
\begin{align}
|B_H\rangle &= p|\Bz\rangle + q|\BzB\rangle, \nonumber \\
|B_L\rangle &= p|\Bz\rangle - q|\BzB\rangle.
\label{eq:mixing}
\end{align}
The time-dependent decay rate for same-sign dileptons is given by
\begin{equation}
  \Gamma_{\ups \to \ell^+ \ell^+}(\dt)
  =\frac{|A_l|^4}{8\taub}e^{-|\dt|/\taub}
  \left|\frac{p}{q}\right|^2
  \left[\cosh \left(\frac{\dg}{2} \dt\right) - \cos \left(\dm \dt\right)\right]
  \label{eq:rates}
\end{equation}
for the $\ell^+ \ell^+$ sample. For the $\ell^- \ell^-$ sample, $p/q$ is replaced by $q/p$. 
Here \dm and \dg are the differences in mass and
decay width between the two mass eigenstates, \taub is
the average lifetime of the two mass eigenstates, \dt is the
proper time difference between the two \B meson decays. In this analysis only
the absolute value of \dt is used.
%%
%It is assumed that 
%$CP$ is conserved in the semileptonic decay of the neutral \B mesons
%and that their amplitudes are equal---i.e., $A_l = \bar{A_l}$.  
It is assumed that the semileptonic decay of the neutral \B meson is
flavour specific and $CP$ conserving, so that $A_l = \bar{A}_l$.
If $CP$ is not conserved in mixing, the condition $|p/q| = 1$ is no
longer required and the decay rates for $\ell^+ \ell^+$ and $\ell^- \ell^-$
samples can differ. 
As can be seen in Eq.~\ref{eq:rates}, the \dt dependence is the
same for the $\ell^+ \ell^+$ and $\ell^- \ell^-$ samples, and therefore 
the $CP$-violation shows up as a \dt-independent charge asymmetry,
defined as
\begin{equation}
A_{sl} \equiv 
\frac{\Gamma_{\ups \to \ell^+ \ell^+} - \Gamma_{\ups \to \ell^- \ell^-}}
     {\Gamma_{\ups \to \ell^+ \ell^+} + \Gamma_{\ups \to \ell^- \ell^-}}
 = \frac{1 - |q/p|^4}{1 + |q/p|^4}
 \simeq \frac{4 Re(\epsilon_B)}{1 + |\epsilon_B|^2}. 
\label{eq:asl}
\end{equation}
Here $\epsilon_B$ corresponds to the $\epsilon_K$ describing
$CP$-violation in the neutral $K$ meson system. Standard Model
calculations give the size of this asymmetry to be of the order of
$10^{-3}$~\cite{cppredict,PRD}.
%assuming that there is no wrong-sign decay ($B^0 \rightarrow X\ell^- \bar\nu$)
%except via \Bz-\BzB mixing.
A significantly larger value would therefore
be an indication of new physics.

Experimentally, measurement of same-sign dilepton events that
originate from $\Bz\Bz$ and $\BzB\BzB$ initial states
requires careful charge-dependent corrections, which must be done in
several steps. First, the contribution from continuum $e^+ e^- \to q\bar q$
(where $q=u,d,s$ or $c$) to
same-sign dilepton events must be
subtracted using off-resonance data. Second, all detected lepton 
tracks must be corrected for charge asymmetries in
the efficiencies for track finding and 
lepton identification, and for 
the probabilities of misidentifying hadrons as leptons.
%misidentification probabilities of
%hadrons to leptons. 
After these corrections, 
the remaining same-sign dilepton events still 
contain backgrounds from $\Bz\BzB$ and $\B^+ \B^-$ events.
The last step of this analysis is to separate the
signal events from these background events using their different behavior
in the \dt distributions. 

\section{Belle Detector}

The data were collected with the Belle detector~\cite{belle} at the KEKB
asymmetric $e^+e^-$ collider~\cite{kekb}.
% between January, 2000 and June, 2002.

The Belle detector is a large-solid-angle magnetic
spectrometer that 
consists of a three-layer silicon vertex detector (SVD),
a 50-layer central drift chamber (CDC) for tracking, a mosaic of
aerogel threshold Cherenkov counters, time-of-flight
scintillation counters (TOF), and an array of CsI(T$\ell$) crystals
for electromagnetic calorimetry (ECL)  located inside of
a superconducting solenoid coil that provides a 1.5~T
magnetic field.  An iron flux-return located outside of
the coil is instrumented to detect $K_L$ mesons and to identify
muons (KLM).  
The integrated luminosity of the data sample is 78 \invfb at the
\ups resonance (``on-resonance'') and 9 \invfb at 60 MeV
below from the \ups resonance (``off-resonance'').

\subsection{Track finding Efficiency}

\begin{figure}[thb]
\includegraphics[width=17cm,clip]{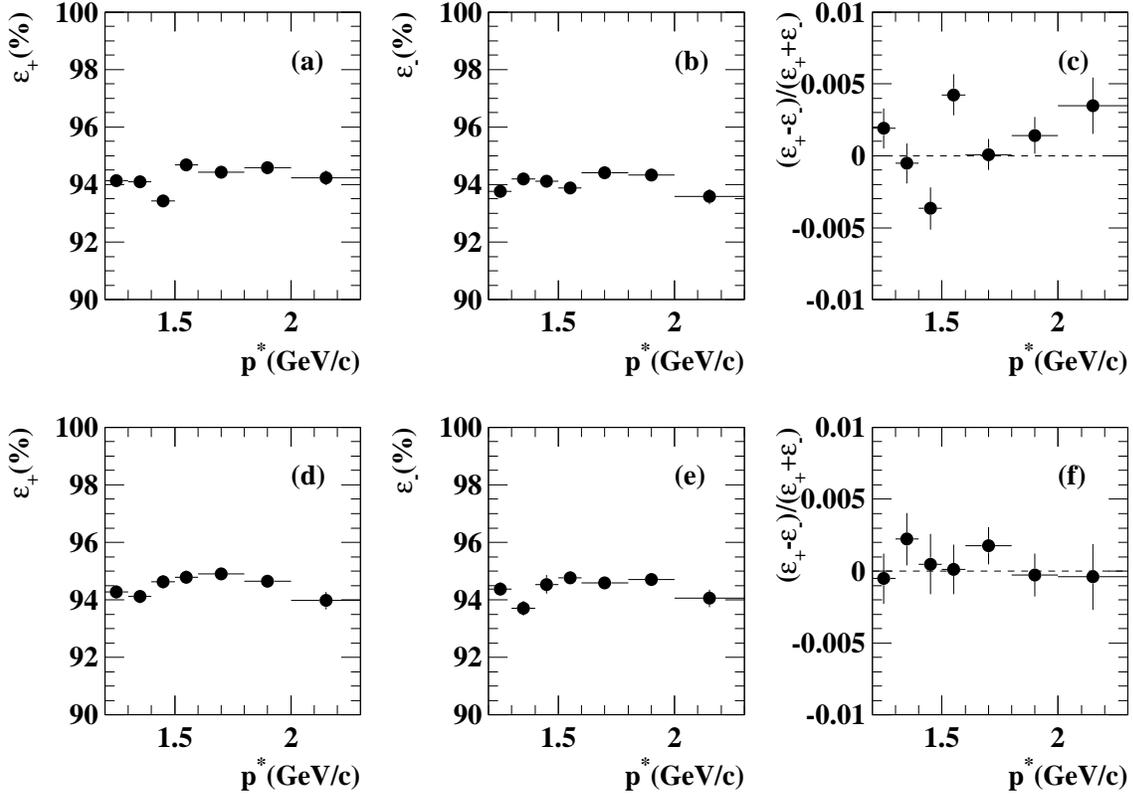}
\caption{Track finding efficiencies as a function of c.m. momentum
 for positron  tracks $\varepsilon_+$ (a), electron tracks
 $\varepsilon_-$ (b), and charge dependence defined as 
 $(\varepsilon_+ - \varepsilon_-)/(\varepsilon_+ + \varepsilon_-)$. 
 Corresponding plots for muon tracks are shown in (d), (e), and (f)}.
\label{seleffe}
\end{figure}

The track finding efficiency is determined by analyzing a sample
where simulated single electron or muon tracks are overlaid on
hadronic events taken from experimental data. Lepton tracks are
generated to cover the region of $1.2~\gevc <p^*<2.3~\gevc$ and
$30^{\circ}<\theta_\text{lab}< 135^{\circ}$, 
where $p^*$ is the lepton momentum in the $e^+ e^-$ center-of-mass (c.m.)
and $\theta_\text{lab}$ is the
angle of lepton track with respect to the $z$-axis in the laboratory
frame. The $z$-axis passes through the nominal interaction point, and
is anti-parallel to the positron beam direction.  Figure~\ref{seleffe}
shows track finding efficiencies for positive and negative tracks
separately and their fractional differences as a function of $p^*$ for
electron and muon tracks. Events in all $\theta_\text{lab.}$ regions
are combined in these plots. The charge dependence of the track finding
efficiency for both electrons and muons is less than 1.0\%.

\subsection{Lepton Identification}
The most important contribution to the electron identification comes
from examination of the ratio of the ECL cluster energy to the track
momentum measured in the CDC. This information is combined with the
shower measurement in the ECL, the specific ionization measurements ($dE/dx$) in
the CDC, and the ACC light yield, to form an electron likelihood
${\cal L}_e$.~\cite{eid}

The two-photon process $e^+e^- \to (e^+e^-)e^+e^-$ is used to 
estimate the electron identification efficiency.
For this data sample, events are required to have: i) 
two tracks with particle ID information inconsistent with a muon hypothesis,
laboratory momenta greater than $0.5~\gevc$ and
transverse momenta greater than $0.25~\gevc$;
ii) at least one ECL cluster with energy greater than $20~\mathrm{MeV}$. 
The two tracks are required to have:
i) an acolinearity angle whose cosine is greater than $-0.997$;
ii) a transverse-momentum sum less than $0.2~\gevc$; and 
    a longitudinal momentum sum of less than $2.5~\gevc$ in the c.m. frame;
iii) an invariant mass less than $5~\gevcc$.  In addition the sum of the 
ECL cluster energies must be between 0.6 \gevc and 6.0 \gevc.
The electron identification efficiency is obtained by taking the ratio of 
the number of tracks selected with the above requirements with and without 
additional electron identification requirements.

For muon identification, CDC tracks are extrapolated to the KLM and 
the measured range and transverse deviation in the KLM is compared with 
the expected values to form a muon likelihood $\mathcal{L}_\mu$~\cite{muonid}.

The muon identification efficiency is determined by analyzing a data sample 
where simulated single-muon tracks are overlaid on the hadronic events 
taken from experimental data. 

\begin{figure}[!thb]
\includegraphics[width=17cm,clip]{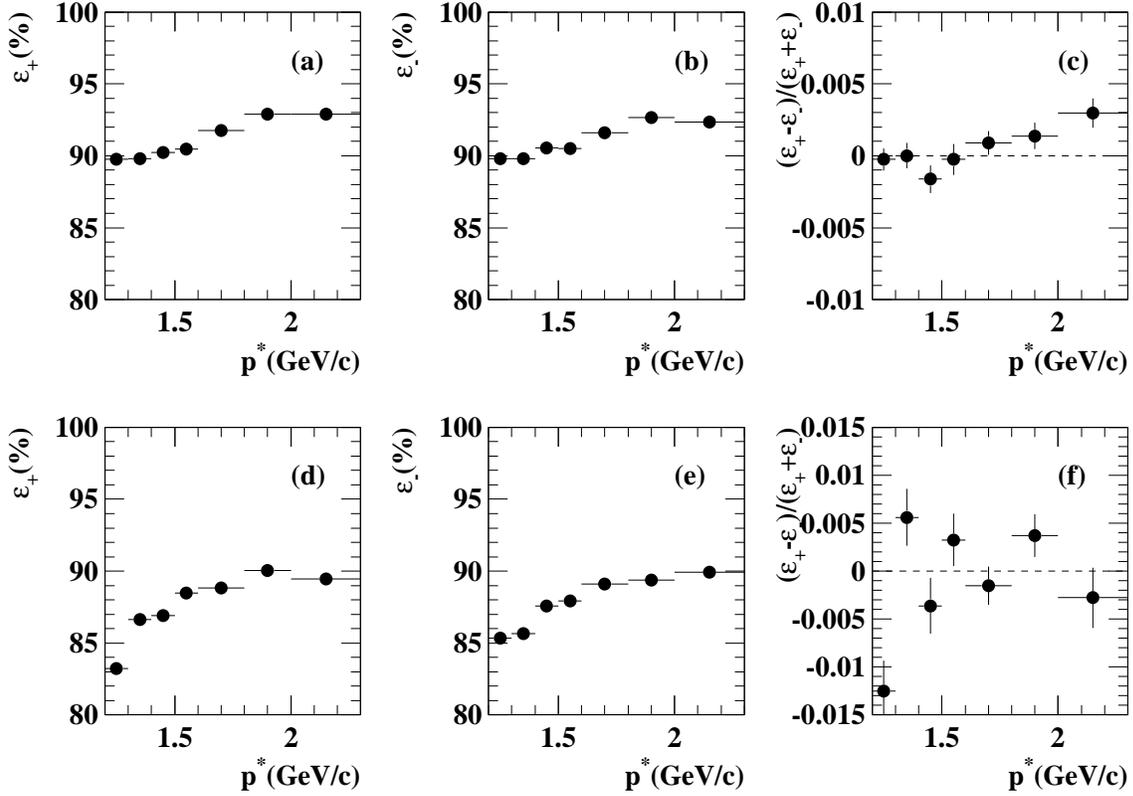}
\caption{Identification efficiencies as a function of c.m. momentum 
for positrons $\varepsilon_+$ (a), electrons $\varepsilon_-$ (b), 
  and charge dependence defined as
 $(\varepsilon_+-\varepsilon_-)/(\varepsilon_+ + \varepsilon_-)$(c). Corresponding plots 
 for muons are shown in (d), (e), and (f). 
}
\label{leptonid}
\end{figure}

Figure~\ref{leptonid} shows the charge-dependent lepton identification
efficiencies, where electron tracks are required to satisfy
$\mathcal{L}_e > 0.8$ and the muon tracks are required to satisfy
$\mathcal{L}_\mu > 0.9$ and the reduced $\chi^2$, of the transverse
deviation in the KLM is required to be less than 3.5. The charge
dependence of both electron and muon identification efficiencies are
less than 1\%.

\subsection{Hadron identification}

Comparison of the ACC light yield to the track momentum, 
the time of flight measurement, and $dE/dx$ 
measurements in the CDC 
are combined to provide hadron likelihoods, $\mathcal{L}_\pi$ for pions, 
$\mathcal{L}_K$ for kaons, and $\mathcal{L}_p$ for protons. 

\subsection{Fake lepton}

The hadron fake rate, which is defined as the probability that a hadron track 
is mis-identified as a lepton, is determined from a sample of 
$K_s \to \pi^+\pi^-$ for pions, 
$\phi \to K^+K^-$ for kaons, and $\Lambda \to p\pi^-$ 
($\bar{\Lambda} \to \bar{p} \pi^+$) for protons.
These decays are selected from a hadronic event sample, which will be described later.
To select these track pair combinations, the closest approach with
respect to the run-dependent interaction point and the position of decay
vertex are used. The $z$ position distance of two tracks and the
deflection angle (except for $\phi \to KK$) at the decay vertex are also used.
%         Ks    Lambda phi
% dr      <0.5  <0.5   <0.5
% dz      <4.0  <4.0   <4.0
% z-dis   <2.0  <2.0   <1.5
% def     <0.1  <0.1   <
% vr      >0.1  >0.1   <0.1
% vz                   <1.5
For each decay, the invariant mass of the two tracks is
calculated after imposing the hadron identification requirement on the
negative (positive) charged track.
The signal yields are obtained by fitting the resulting mass distributions to sums of
double Gaussian signal terms and smooth background functions in two ways:
once without imposing any particle identification requirement and
again after imposing the lepton identification requirement on the positive
(negative) charged track.  The ratios of the two signal yields give
the fake rates for the positive (negative) charged tracks.
The following cuts are placed on the likelihood ratios:
$\mathcal{L}_\pi/(\mathcal{L}_\pi + \mathcal{L}_K)> 0.8$ for pions in 
$K_S \to \pi^+ \pi^-$, 
$\mathcal{L}_K/(\mathcal{L}_K+ \mathcal{L}_\pi)> 0.8$ for kaons in 
$\phi \to K^+ K^-$, and 
$\mathcal{L}_\pi/(\mathcal{L}_\pi + \mathcal{L}_p)> 0.8$ for pions in 
$\Lambda \to p \pi^-$.
The rate of pions faking electrons is at most 0.1\% for both charges
and shows no significant charge dependence. The rate of kaons faking
electrons decreases rapidly as $p_\text{lab}$ becomes larger and is
less than 0.2\% for $p_\text{lab} > 1.4~\gevc$ with no significant
charge dependence. While the rate of protons faking electrons is
nearly zero, the rate for anti-protons faking electrons is as large as 4\% due to the large
anti-proton annihilation cross section in the ECL. The rate of pions
faking muons is about 1\% for $p_\text{lab} > 1.5~\gevc$ and shows
no significant charge dependence. The rate of kaons faking muons is 1\%
to 2\% and that for $K^+$ is about 50\% larger than $K^-$ due to the
larger kaon-nucleon cross section for the $K^-$. The rate of protons faking
muons is less than 0.4\% and shows no clear charge dependence.

\begin{figure}[!thb]
\includegraphics[width=17cm,clip]{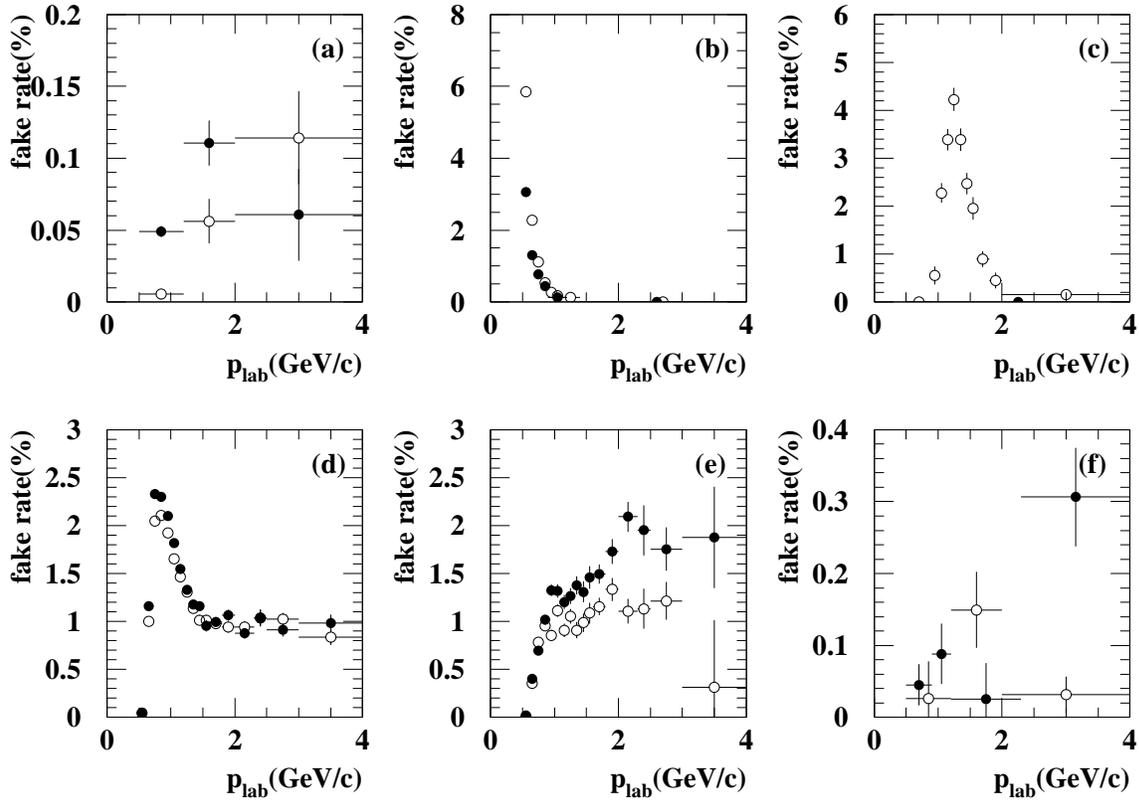}
%\includegraphics[width=8.5cm,clip]{../fake_rate/fake_rate_pi_eidcomb.eps}
%\includegraphics[width=8.5cm,clip]{../fake_rate/fake_rate_pi_muid.eps}
%\vspace{12cm}
\caption{
  Rates of pions ((a) and (d)), kaons ((b) and (e)), and protons ((c)
  and (f)) faking electrons and muons vs laboratory momentum.  Filled
  circles are for positive tracks and open circles are for negative tracks.
  The increase in the rate of kaons faking electrons at low momentum
  clearly visible in (b) is due to the overlap of the electron and
  kaon energy loss bands.}
\label{fake}
\end{figure}

\section{Event Selection}

\subsection{Hadronic Event Selection}

Hadronic events are required to have at least five tracks, an event
vertex with radial and $z$ coordinates within $1.5~\cm$ and $3.5~\cm$
respectively of the nominal beam interaction point, a total
reconstructed c.m. energy greater than $0.5~W$ ($W$ is the \ups c.m.
energy), a net reconstructed c.m. momentum with a $z$ component less
than 0.3~$W$$/c$, a total energy deposited to ECL between $0.025$ and
$0.9~W$, and a ratio $R_2$ of the second and zeroth Fox-Wolfram
moments~\cite{Fox-Wolfram} less than 0.7.

\subsection{Dilepton Event Selection}

Lepton candidates are selected from the charged tracks by requiring
$\mathcal{L}_e >0.8$ for electrons or $\mathcal{L}_\mu > 0.9$ and a reduced
$\chi^2$ of the transverse deviation in the KLM of less than 3.5 for muons. 
In both cases a distance of closest approach to the run-dependent interaction
point less than $0.05~\cm$ radially and $2.0~\cm$ in $z$ is required.  At least
one SVD hit per track in the $r$-$\phi$ view and two SVD hits in the $r$-$z$ view is required.
To eliminate electrons from $\gamma \to e^+ e^-$
conversions, electron candidates are paired with all other oppositely
charged tracks and the invariant mass (assuming the electron mass
hypothesis) $M_{e^+e^-}$ is calculated. If $M_{e^+e^-} < 100~\mevcc$,
the electron candidate is rejected. If a hadronic event contains more
than two lepton candidates, the two with the highest c.m. momenta are
used.

The two lepton candidates must satisfy additional criteria.  
The c.m. momentum of each lepton is required to be in the range 
$1.2~\gevc < p^* < 2.3~\gevc$. The lower cut reduces contributions from
secondary charm decay. The upper cut reduces continuum
contributions. 
Each lepton track must satisfy $30^\circ
<\theta_\text{lab}<135^\circ$. This cut selects tracks with better
$z$ vertex resolution and better lepton identification.
Events that contain one or more \jpsi candidates are rejected.
The invariant mass of each candidate lepton paired with each
oppositely charged track (assuming the correct lepton mass hypothesis)
is calculated. If the invariant mass falls into the \jpsi region,
defined as $-0.15~\gevcc < ( M_{e^+ e^-} - M_{\jpsi})< 0.05~\gevcc$ or 
$-0.05~\gevcc < (M_{\mu^+\mu^-} - M_{\jpsi})< 0.05~\gevcc$,
the candidate event is rejected. The looser lower cut for
the electron pair invariant mass is used to reject \jpsi decays
with a low invariant
mass due to bremsstrahlung of the daughter electron(s).

\begin{figure}[!thb]
\includegraphics[width=17cm,clip]{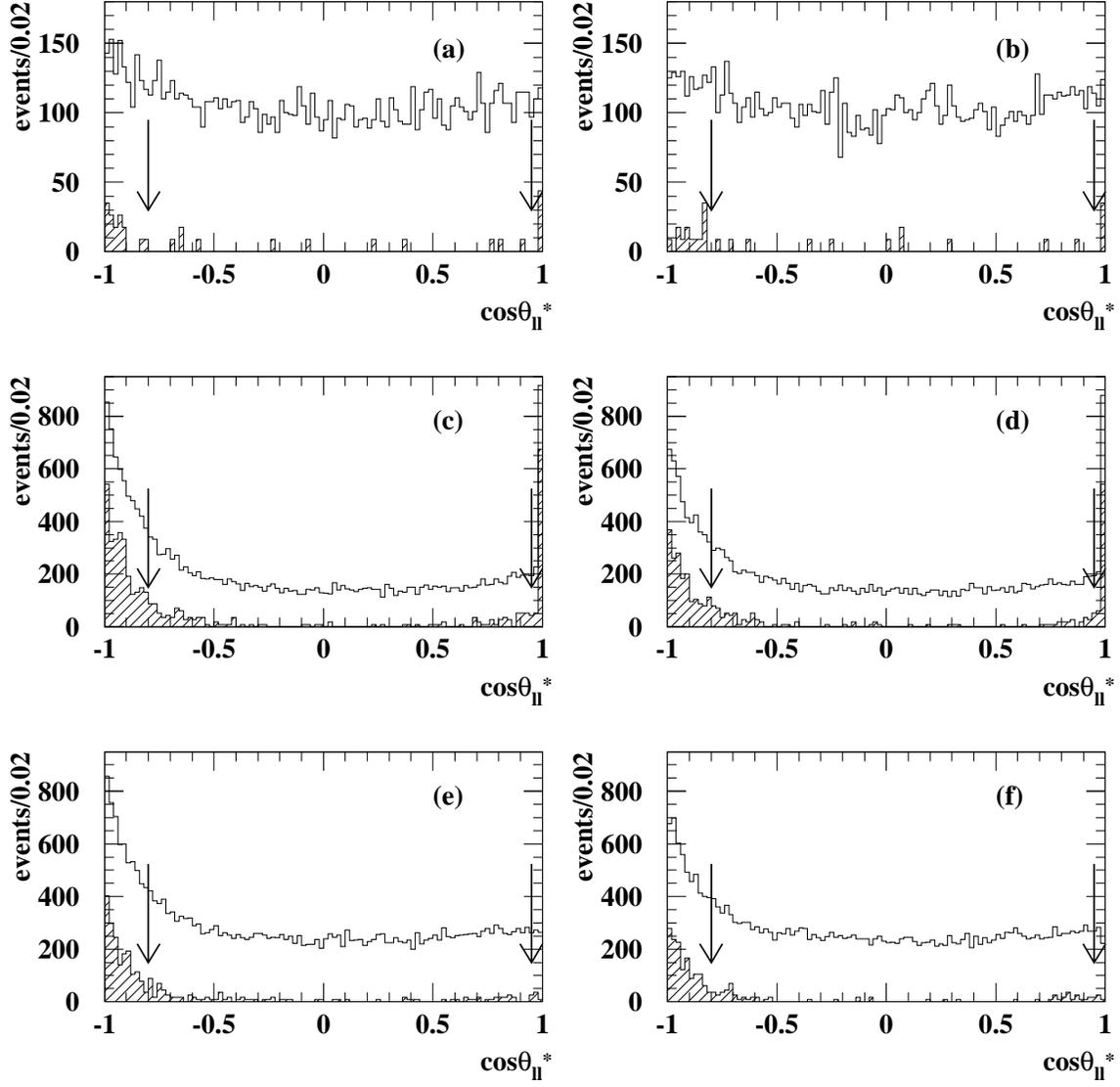}
\caption{
$\cos \theta^*_{\ell\ell}$ distributions for the dilepton
samples of the on-resonance (open histogram) and scaled
off-resonance (filled histogram) data.
(a),(b) show $ee$ events, (c),(d) show $\mu\mu$ events and (e),(f) are
from $e\mu$ combinations. (a),(c) and (e) are the $++$ charge case and
(b), (d) and (f) are the $--$ charge case.
The arrows indicate the selected range.
}
\label{openang}
\end{figure}

As can be seen in Fig.~\ref{openang}, distributions of the opening
angle of the two tracks in the c.m. frame, $\cos \theta^*_{\ell\ell}$,
for the $\mu \mu$ and $e\mu$ pairs show distinct peaks in the 
back-to-back direction ($\cos\theta^*_{\ell\ell}\simeq -1$). This
background
% in the muon candidates 
is caused by jet-like continuum events and events with a
primary lepton and a secondary lepton originating from the same \B meson.
Also spikes can be seen at $\cos\theta^*_{\ell\ell} = 1$ in the $\mu\mu$
pairs. 
This structure is caused by jet-like continuum events where a non-muon
track is identified as muon because it is assigned hits in the KLM from a
true muon.
%This background is caused by non-muon track in jet-like cotinuum event
%is accidentaly identified as muon because of sharing hit points in the KLM
%with true muon.
The opening dilepton angle in the
c.m. frame $\theta^*_{\ell\ell}$ is required to satisfy 
$-0.80 < \cos\theta^*_{\ell\ell} < 0.95$ in order to reduce this background.

With these selection criteria there are 46551 positive and 45507
negative same sign dilepton events found in the on-resonance data.
Continuum contributions are estimated to be 2229.8 for positive and
1556.5 for negative same sign events, based on the yield from
off-resonance data. To estimate the continuum contribution from off-resonance
data, off-resonance yields were scaled by the integrated luminosities and
cross-section ratio. The scaling factor is defined in the Eq.~\ref{eq:bbyield}.
%These numbers, broken down 
These dilepton yields decomposed into different lepton
categories, are given in Table~\ref{num}.

\begin{table}[!htb]
  \begin{center}
    \caption{Number of dilepton events}
    \begin{tabular}{|c|r|r||r|r||r|r|}
      \hline
      \hline
      & \multicolumn{2}{c||}{on-resonance}&\multicolumn{2}{c||}{off-resonance}&\multicolumn{2}{c|}{continuum}\\
      \cline{2-7}
      combination & positive & negative & positive & negative & positive & negative \\
      \hline
      $ee$        &  9059$\pm$ 95.2 &  9028$\pm$ 95.0 &  11$\pm$ 3.3 &  11$\pm$ 3.3 &   96.2$\pm$ 28.9 &  96.2$\pm$ 28.9 \\
      $\mu\mu$    & 14672$\pm$121.1 & 14014$\pm$118.4 & 144$\pm$12.0 & 100$\pm$10.0 & 1259.2$\pm$104.9 & 874.4$\pm$ 87.4 \\
      $e\mu$      & 22802$\pm$151.0 & 22453$\pm$149.8 & 100$\pm$10.0 &  69$\pm$ 8.3 &  874.4$\pm$ 87.4 & 603.4$\pm$ 72.6 \\
      \hline
      total       & 46533$\pm$215.7 & 45477$\pm$213.3 & 255$\pm$16.0 & 180$\pm$13.4 & 2229.8$\pm$139.6 &1574.0$\pm$117.3 \\
      \hline
      \hline
    \end{tabular}
    \label{num}
  \end{center}
\end{table}

\subsection{\dz Determination}

The $z$-coordinate of each \B meson decay vertex is the production 
point of the daughter lepton, which is determined from 
the intersection of the lepton track with the run-dependent profile of 
the interaction point. 
$ |\Delta z| =  |z(\ell_1) - z(\ell_2)|$ is the distance between 
the $z$-coordinates of the two leptons. 

In order to estimate the detector resolution in the \dz determination
\jpsi decays to $e^+e^-$ and $\mu^+\mu^-$ are used. In these events
the two tracks originate from the same point, so the measured \dz,
after the background contribution is subtracted, yields the detector
resolution. Candidate \jpsi are selected using the same requirements
as the dilepton events except the \jpsi veto.  The \jpsi signal regions
are defined as $3.0~\gevcc < M(e^+e^-) < 3.14~\gevcc$ and $3.05~\gevcc
< M(\mu^+\mu^-) < 3.14~\gevcc$ and the sideband region as $3.18~\gevcc
< M(\ell^+\ell^-) < 3.50~\gevcc$ for both electrons and muons.

The invariant mass distributions of $J/\psi$ candidates are fitted to a
function given by
\begin{equation}
N(M) = h_0 e^{ -\frac{(M - M_0)^2}{2S^2 } }
+ h_1 e^{ -\frac{(M - M_0)^2}{2 {\sigma_1}^2} }
+ A(M-B)^2 + C.
\end{equation}
Here $h_0$ and $h_1$ are the heights of both Gaussians, $M_0$ is
the Gaussian mean which is common to two Gaussians,
$\sigma_0$ and $\sigma_1$ are the width of the Gaussians.
A parameter $S$, given as $S = \sigma_0$ for $M \ge M_0$ and
$S = \sigma_0 + \alpha(M - M_0)$ for $M < M_0$,
is introduced to modify the lower mass tail of one of the Gaussians for
the effect of bremsstrahlung using another parameter $\alpha$.
$A$, $B$ and $C$ are the parameters of the background function.
The
\dz distribution of the sideband region is scaled to the background
yield in the signal region and subtracted from the signal region \dz
distribution.

\begin{figure}[!thb]
\includegraphics[width=17cm,clip]{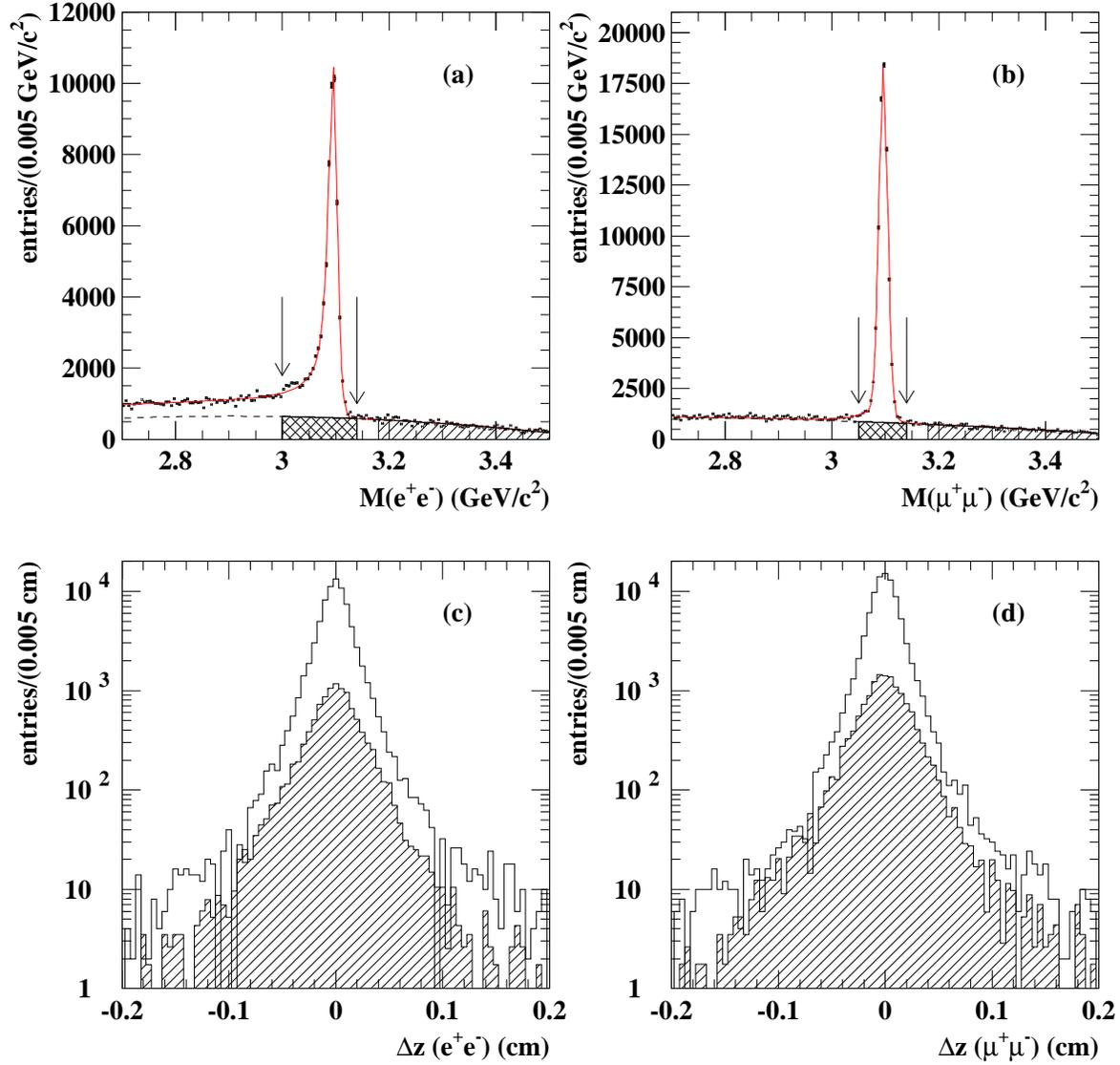}
%\vspace{12cm}
\caption{Mass distributions for $\jpsi \to e^+e^-$ (a) and $\jpsi \to 
  \mu^+\mu^-$ (b). The arrows indicate the signal region for each
  decay mode.  The dashed lines indicate the fitted background
  component and the solid lines show the total fit results.  The
  cross hatched area shows the estimated combinatorial background in
  the signal region and single hatched area is the sideband region
  used to estimate background \dz distribution. The \dz distributions
  for $\jpsi \to e^+e^-$ (c) and $\jpsi \to \mu^+\mu^-$ (d). Open
  histograms are for all \jpsi candidates in the signal region and
  hatched histograms are for the background.}
\label{mrjpsi}
\end{figure}

The \jpsi mass distributions and the \dz distributions are 
shown in Fig.~\ref{mrjpsi}.
The RMS of the \dz distributions are 193 $\micron$ for
$\jpsi \to e^+ e^-$, $177~\micron$
for $\jpsi \to \mu^+ \mu^-$, and $185~\micron$ for the combined
$\jpsi \to \ell^+\ell^-$.

\subsection{Subtraction of Continuum Events}
A sample of dilepton events originating from $B\bar{B}$ events is obtained by
subtracting the luminosity and cross-section scaled off-resonance 
data from the on-resonance data. 
Since the kinematics of dilepton candidates are generally different
in these two data samples, the subtraction should, in principle be 
performed in a  six dimensional 
($p^*_1$, $p^*_2$, $\theta^*_1$, $\theta^*_2$, $\theta^*_{\ell\ell}$, \dz)
space for each lepton flavour and charge combination,
where $\theta^*_{1(2)}$ is the polar angle with respect to the beam axis
of more(less) energetic lepton
in c.m. frame. 
The number of $B\bar{B}$ events is obtained from
\begin{equation}
N_{\B \bar{\B}}(p^*_1, p^*_2, \theta^*_1, \theta^*_2,\theta^*_{\ell\ell}, \dz)=
N_{\text{on}}(p^*_1, p^*_2, \theta^*_1, \theta^*_2,\theta^*_{\ell\ell}, \dz)- 
\frac{\int \mathcal{L}_{\text{on}} dt}
     {\int \mathcal{L}_{\text{off}} dt}
\frac{s_{\text{off}}}
     {s_{\text{on}}} 
     N_{\text{off}}
   ( p^*_1, p^*_2, \theta^*_1, \theta^*_2, \theta^*_{\ell\ell}, \dz )
\label{eq:bbyield}
\end{equation}
where $N_{B \bar B}$ and $N_{\rm{on}(\rm{off})}$ are 
the dilepton yields of $B \bar B$ origin and  on(off)-resonance data,  
respectively,  
$\int \mathcal{L}_{\rm{on}(\rm{off})} dt$ and $s_{\rm{on}(\rm{off})}$ are 
the integrated luminosities and the square of c.m. energies for
on(off)-resonance, respectively. 
$\theta^*_{\ell\ell}$ is included here because this variable behaves 
distinctly differently in the two data samples for the cases containing
muons, as shown in Fig.~\ref{openang}. 

\begin{figure}[!thb]
%\vspace{12cm}
\includegraphics[width=17cm,clip]{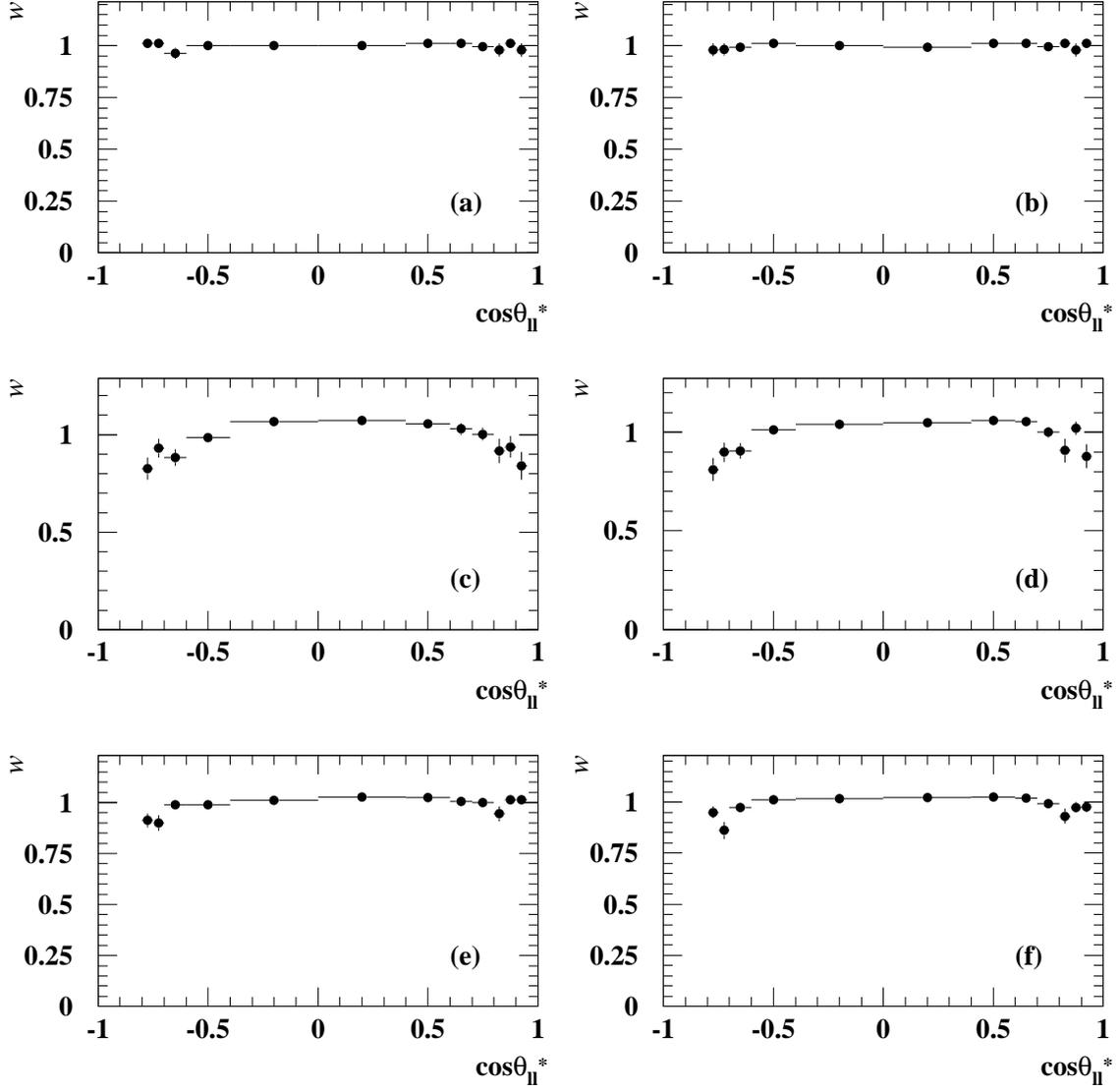}
 \caption{$\cos \theta^*_{\ell\ell}$ dependence of the weighting factor for the 
 fraction of the dilepton yield of $B \bar B$ origin in the on-resonance 
 data for $e^+e^+$ (a) and $e^-e^-$ (b) and
corresponding for $\mu\mu$ (c) and (d) and $e\mu$ (e) and (f).}
\label{wopang}
\end{figure}

Given the available statistics, this approach is not possible. Instead, we perform
the subtraction by weighting the on-resonance and off-resonance yields
for one of the six kinematical variables, while integrating over the five other variables.
We obtain weighting factors for the six variables by
repeating this procedure. Since, t oa first approximation, the six
variables are not correlated with each other, this approach provides
the $B \bar B$ yield in the six variable space.
The weighting factors  are given by
$w(k) =(1/r_{BB})
  (N_{\mathrm{on}}(k) -f N_{\mathrm{off}}(k))/N_{\mathrm{on}}(k)$
where $k$ denotes each of six variables,
$f$ is the scaling factor for the luminosity and c.m. energy introduced
in Eq.~\ref{eq:bbyield},
and $r_{BB} \equiv N_{B \bar B}^\mathrm{total}/ N_\mathrm{on}^\mathrm{total}$
is the fraction of total $B \bar B$ in the on-resonance yield after integrtating over
all six variables and is used for the normalization.
While the weighting
factors show very little dependence on $p^*_1$, $p^*_2$, $\theta^*_1$,
and $\theta^*_2$ for all combinations of lepton flavours and charges,
a clear dependence is observed for $\theta^*_{\ell\ell}$ in the case
of $\mu \mu$ and $e\mu$ data samples as shown in Fig.~\ref{wopang}.
A clear dependence on $\Delta Z$ is also seen for all lepton pair
combinations.
%\dz also shows the clear dependence for all combinations of lepton types.

Using this method, the dilepton candidate yield for each 
lepton flavour and charge combination is then given 
in terms of the on-resonance yield and the weighting factors
by 

\begin{equation}
N_{B\bar{B}}(p^*_1, p^*_2, \theta^*_1, \theta^*_2, \theta^*_{\ell\ell}, \dz ) 
= r_{BB}  \prod_{k} w(k) 
N_{\rm{on}}( p^*_1, p^*_2, \theta^*_1, \theta^*_2, \theta^*_{\ell\ell}, \dz ).
\label{eq:nbbbar}
\end{equation}
The \dz dependence of the dilepton yields
are obtained by projecting
$N_{B\bar{B}}(p^*_1, p^*_2, \theta^*_1, \theta^*_2, \theta^*_{\ell\ell}, \dz)$
onto the $|\dz|$ axis.
\section{Result}

\subsection{Corrections to lepton candidates}

\begin{figure}[!thb]
\includegraphics[width=17cm,clip]{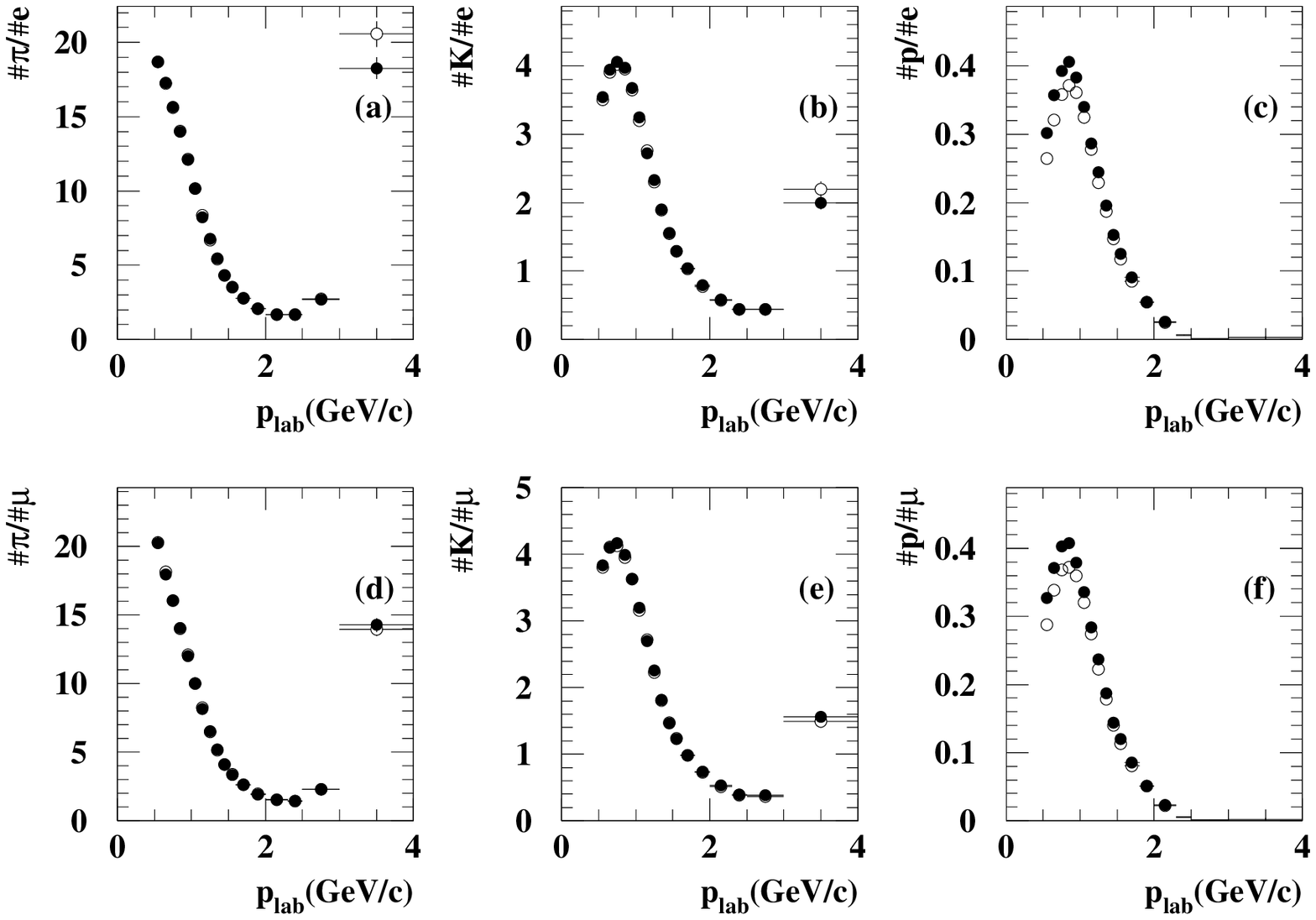}
%\vspace{12cm}
\caption{Relative multiplicities of hadrons as a function of $p_\text{lab}$
  with respect to leptons. The hadron and lepton species are indicated
  on the vertical axis of each graph.  Filled circles are for positive
  tracks and open circles are for negative tracks.  Though in (a),
  (b), (d) and (e), the difference between positive and negative is
  less than 1\%, in both (c) and (f), proton rate is larger than
  anti-proton rate by about 8\%.}
\label{hlratio}
\end{figure}

%The number of detected lepton pairs for each lepton flavour and charge
%combination $N^\pm_{\text{det}}$ is related the number of true leptons 
%$N^\pm_\ell$ by
The number of detected leptons for each lepton flavour and charge
$N^\pm_{\text{det}}$ is related to the number of true leptons 
$N^\pm_\ell$ by
\begin{eqnarray}
%  N^\pm_{\text{det}}(p^*,\theta_{\text{lab}}) &=& 
%  N^\pm_\ell(p^*,\theta_{\text{lab}})
%  \varepsilon^\pm_{\text{trk}}(p^*,\theta_{\text{lab}})
%  \{\varepsilon^\pm_{\text{pid}}(p^*,\theta_{\text{lab}})
%  + r^\pm_{\pi\ell}(p^*,\theta_{\text{lab}})\eta^\pm_{\pi\ell}(p^*,\theta_{\text{lab}}) 
%  \nonumber \\
%  & &+ r^\pm_{K\ell}(p^*,\theta_{\text{lab}})\eta^\pm_{K\ell}(p^*,\theta_{\text{lab}})
%  + r^\pm_{p\ell}(p^*,\theta_{\text{lab}})\eta^\pm_{p\ell}(p^*,\theta_{\text{lab}}) \},
  N^\pm_{\text{det}}(p^*,\theta_{\text{lab}}) =
  N^\pm_\ell(p^*,\theta_{\text{lab}})
  \varepsilon^\pm_{\text{trk}}(p^*,\theta_{\text{lab}})
  \{\varepsilon^\pm_{\text{pid}}(p^*,\theta_{\text{lab}})
  + \sum_{h=\pi,K,p} r^\pm_{h\ell}(p^*,\theta_{\text{lab}})\eta^\pm_{h\ell}(p^*,\theta_{\text{lab}})\},
\label{ndet}
\end{eqnarray}
where $\varepsilon^\pm_{\text{trk}}$ and $\varepsilon^\pm_{\text{pid}}$ are
the efficiencies for track finding and lepton identification, 
$r^\pm_{h \ell}$ is the relative multiplicity of hadron 
$h$ with respect to lepton $\ell$ in the $B \bar B$ event, 
and $\eta^\pm_{h \ell}$ is the rate of hadrons $h$ faking  leptons $\ell$. 
The relative multiplicities are determined from $\Bz\BzB$ 
Monte Carlo (MC) events, and are shown in Fig.~\ref{hlratio}.

% pi/e+ 9.84870 +- 0.00799804
% pi/e- 9.86606 +- 0.00799256
% K/e+  2.62686 +- 0.0023883
% K/e-  2.60991 +- 0.0023694
% p/e+  0.250171 +- 0.000432722
% p/e-  0.231581 +- 0.00041225

% pi/mu+ 9.69384 +- 0.00781587
% pi/mu- 9.72890 +- 0.00783158
% K/mu+  2.58556 +- 0.00233732
% K/mu-  2.57362 +- 0.0023247
% p/mu+  0.246237 +- 0.000425247
% p/mu-  0.228361 +- 0.000405988

% pi/e+ 9.8487 +- 0.0080
% pi/e- 9.8661 +- 0.0080
% K/e+  2.6269 +- 0.0024
% K/e-  2.6099 +- 0.0024
% p/e+  0.25017 +- 0.00043
% p/e-  0.23158 +- 0.00041

% pi/mu+ 9.6938 +- 0.0078
% pi/mu- 9.7289 +- 0.0078
% K/mu+  2.5856 +- 0.0023
% K/mu-  2.5736 +- 0.0023
% p/mu+  0.24624 +- 0.00043
% p/mu-  0.22836 +- 0.00041

Using the measured efficiencies and fake rates and the MC determined relative
multiplicities, the correction factors
$N_\ell/N_{\text{det}}$ were determined in 7 bins of $p^*/\gevc$   
(1.2--1.3, 1.3--1.4, 1.4--1.5, 1.5--1.6, 1.6--1.8, 1.8--2.0 and 2.0--2.3),
and 
8 bins of $\theta_{\text{lab}}$ 
(30$^\circ$--37$^\circ$, 37$^\circ$--50$^\circ$, 50$^\circ$--77$^\circ$,
77$^\circ$--82$^\circ$, 82$^\circ$--111$^\circ$, 111$^\circ$--119$^\circ$,
119$^\circ$--128$^\circ$ and 128$^\circ$--135$^\circ$). 
The fake rates are measured in the laboratory frame. 
For the correction, they are converted into quantities in 
($p^*$, $\theta_{\text{lab}}$). 

After the correction, the dilepton sample contains true leptons that come 
either from prompt neutral \B meson decay (signal) or from background
processes such as charged \B meson decay, secondary charm decay, or
other leptonic \B meson processes.

\subsection{Fit to \dz Distribution}
A binned maximum likelihood fit with signal and background
contributions is used to extract $\Asl(|\dz|)$ from the \dz
distribution. The \dz distribution for the signal events is given by 
Eq.~\ref{eq:rates} assuming $\dg$ is negligible~\cite{PDG} as,
\begin{equation}
  P^{\text{SS}} \propto e^{-|\dt|/\taub}(1-\cos(\dm \dt)),
  \label{pmix}
\end{equation}
convolved with the detector response function described earlier. 
Here \taub is the \Bz lifetime and \dm is the mixing parameter.
These parameters are fixed to their world average values~\cite{PDG}.
%%Those values are
%%$1.542 \pm 0.016$ ps and $0.489 \pm 0.008$ ps$^{-1}$, respectively.

\begin{figure}[!thb]
\includegraphics[width=17cm,clip]{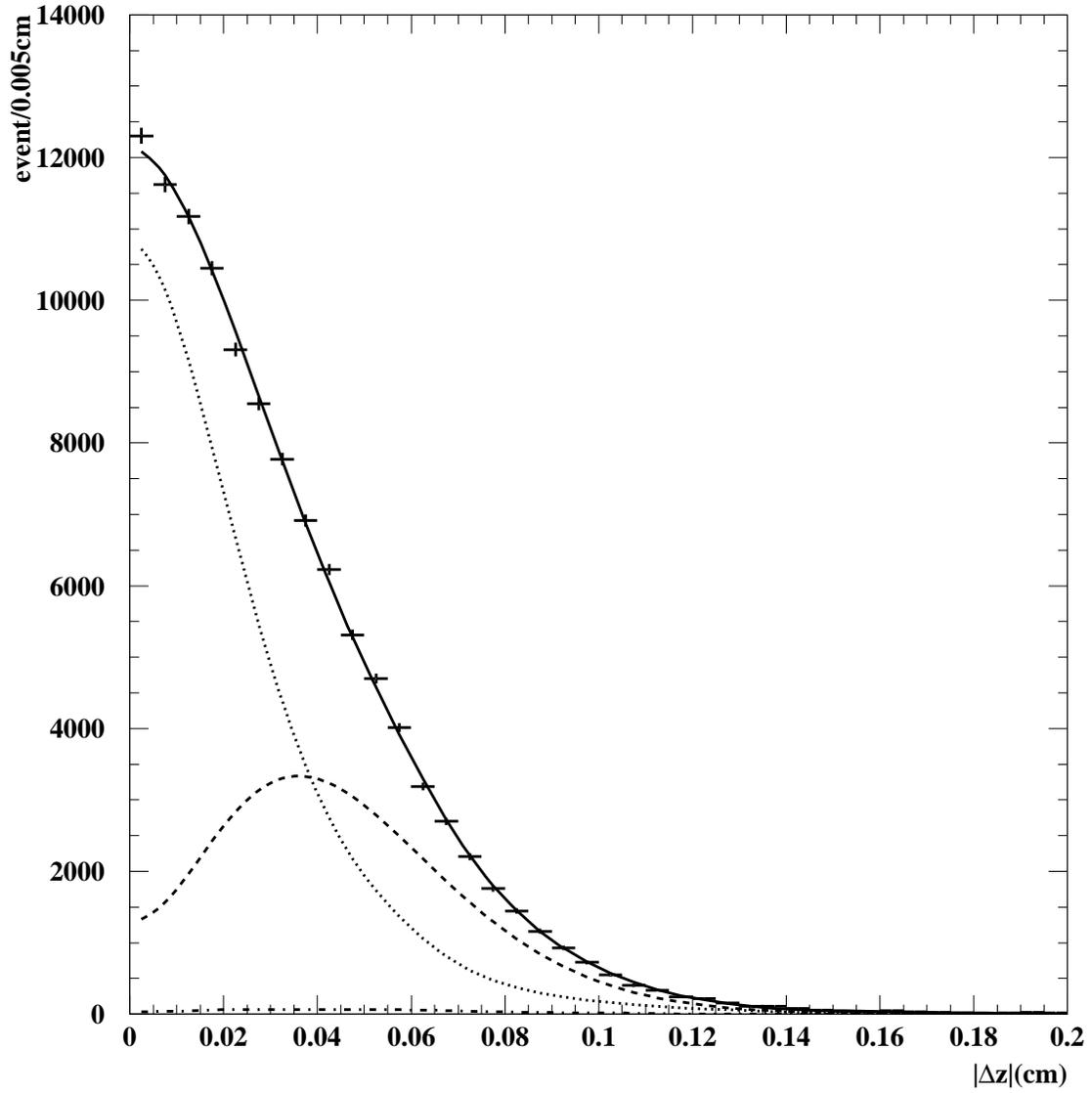}
 \caption{\dz distribution for the true dilepton
   events($++$ and $--$ are combined). 
   Points with error bars are data. The dot-dashed line shows the
   contribution from \CT backgrounds, the dotted line shows the \WT
   background contributions, the dashed line indicates the signal
   component and the solid line indicates the total of the fit.}
\label{dz}
\end{figure}

The backgrounds are placed into two categories; correctly tagged
(\CT), and wrongly tagged (\WT).
The \CT category mainly contains events in which both leptons
come from secondary charm decay in
$\Bz\BzB \to \Bz\Bz
(\BzB\BzB)$ (mixed) processes. The \WT category contains events in
which one lepton is from secondary charm decay of unmixed $\Bz\BzB$ or
$B^+ B^-$
and the other is from a semi-leptonic $B$ decay.
Though background \dz distributions are estimated using MC
simulations, this MC background \dz distribution overestimates the
\dz resolution in the data.
To correct for this the MC \dz
distribution is convolved with a $\sigma = (68 \pm 19)~\micron$
Gaussian~\cite{hastings}.
The \dz distributions for the true same-sign dilepton events
where positive($++$) and negative($--$) samples are combined,
are shown 
in Fig.~\ref{dz} together with the fit results. 
The $\chi^2/n.d.f.$ of the fit is 72.17/38.
In the fit, the ratio of \CT to \WT is fixed at the MC value, and
only the ratio of signal and background is allowed to float.
The MC estimated \CT and \WT
contributions to the \dz distribution are shown in Fig.~\ref{dz}.

\subsection{Charge Asymmetry}

The measured same-sign dilepton charge asymmetry is defined as
\begin{equation}
A_{\ell\ell}(\Delta z) = \frac{N^{++}(\Delta z) - N^{--}(\Delta z)}{N^{++}(\Delta z) + N^{--}(\Delta z)},
\end{equation}
where $N^{\pm\pm}(\Delta z)$ are the $\Delta z$ distributions of the 
true dilepton yields. 

\begin{figure}[!thb]
\includegraphics[width=17cm,clip]{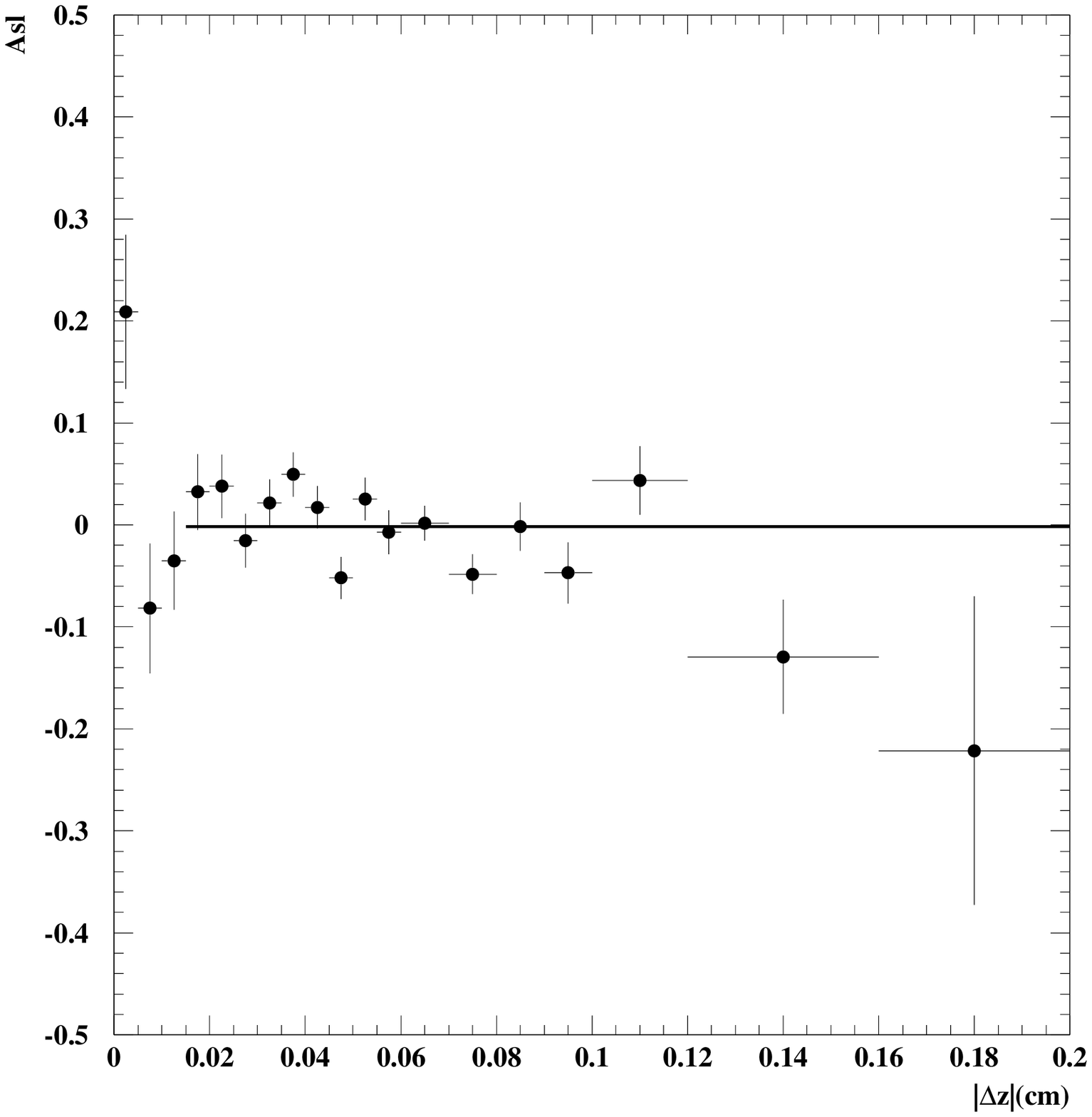}
\caption{$|\dz|$ distribution for \Asl.}
\label{asl}
\end{figure}

Since $N^{\pm\pm}(\Delta z)$ are the sum of signal and background,
$N^{\pm\pm}(\Delta z) = N_s^{\pm\pm}(\Delta z) + N_b^{\pm\pm}(\Delta z)$,
the dilepton charge asymmetry  $\Asl$ is related to $A_{\ell \ell}$
by 

\begin{equation}
A_{\ell\ell}(\Delta z) 
    = \frac{N_s^{++}(\dz) - N_s^{--}(\dz)}{N_s(\dz)} 
      \frac{N_s (\dz)}{( N_s (\dz)+ N_b (\dz))} 
    = A_{sl}(\dz) d(\dz)
\label{fasl}
\end{equation}
where $N_s = N_s^{++} + N_s^{--}$ and $N_b = N_b^{++} + N_b^{--}$. 
A dilution factor, $d(\dz) = N_s (\dz)/ (N_s (\dz) + N_b(\dz))$, is
calculated using the signal and background yields, which are determined
in the fit given in Fig.~\ref{dz}. The result of $\Asl(\dz)$ which is
determined from measured $A_{\ell\ell}(\Delta z)$ and dilution factor
$d(\Delta z)$ is shown in Fig.~\ref{asl}.

The dilepton charge asymmetry is $\Asl(|\dz|)$ is a time integrated
quantity and does not depend on \dz.
Fitting this distribution to a constant in the region of
$0.015~\cm<|\Delta z|<0.200~\cm$, yields $\Asl = (-0.13 \pm 0.60)\%$ and
the $\chi^2/n.d.f.$ is 68.73/36.
The optimum fitting range is determined using a MC study.

\subsection{Cross Checks}

As a consistency check, \Asl is obtained for the $ee$, $\mu \mu$, and $e
\mu$ data samples, separately. The results, $\Asl(ee) = (-1.41 \pm
1.13)\%$, $A_{sl}(\mu \mu) = (+1.66 \pm 1.30)\%$, and $A_{sl}(e\mu)
=(-0.44 \pm 0.94)\%$, are consistent with the primary result.
Here the errors are statistical only.  

In the extraction of \Asl from $A_{\ell\ell}(\dz)$ using 
Eq.~\ref{fasl}, it is assumed that $N_b^{++} = N_b^{--}$. The validity of
this assumption is confirmed by repeating the fit without it. This yields
$N_b^{++} = 20452 \pm 126$ and $N_b^{--} = 20028 \pm
124$ in the range $0.015~\cm < |\dz| < 0.200~\cm$, which is consistent
with the initial assumption.

\subsection{Systematic errors}

Systematic errors in the determination of \Asl come from uncertainties
in: i) the event selection criteria, ii) corrections for efficiencies
of track finding and lepton identification and for lepton
misidentification, iii) the continuum subtraction, iv) the \dz fit for the
dilepton sample, v) the \dz fit for the determination of \Asl. 

Uncertainties in the event selection are estimated by repeating the
analysis with varied cut values. For the track selection, the
$\theta_\text{lab}$ cut is varied from the nominal
$30^{\circ}$--$135^{\circ}$ to $51^{\circ}$--$117^{\circ}$ (barrel
detector part only) in several steps, the closest-approach cut in
$r \phi$ from its nominal value of $0.05~\cm$ to $0.02~\cm$ in six steps, the
closest-approach cut in $z$ from $2~\cm$ to $1~\cm$ in five steps, the
$p^*$ cut from its nominal value of $1.2~\gevc$ to $1.3~\gevc$ in four steps, the
$\cos\theta^*_{\ell\ell}$ cut from nominal
$-0.80 <\cos\theta^*_{\ell\ell}< 0.95$
to $-0.70 <\cos\theta^*_{\ell\ell}<0.90$
in several steps, the requirement on the number of SVD hits by $\pm 1$
from nominal value of greater or equal to two for $z$ and one for
$r \phi$.  In addition, the mass cuts that reject \jpsi and $\gamma \to
e^+ e^-$ are widened by 20\% and 50\%.

To estimate the systematic error from the continuum subtraction, the
analysis is repeated with the continuum subtraction varied by the
statistical error of the off-resonance yield. 

The contributions from track corrections are
estimated by varying each of the efficiencies, fake rates, and
relative multiplicities by $\pm 1\sigma$.

The contribution from the detector \dz response function is estimated
by changing the response function width according to the statistics of \dz
distribution of $\jpsi \to \ell^+\ell^-$ sample.

The contribution from the $68~\micron$ smearing is estimated by
repeating the analysis with $50~\micron$ and $87~\micron$ smearing, 
which are the values obtained when the $\chi^2$ for the \dz fit is changed by one compared with the
default fit. The contributions from uncertainties in \dm and \taub are also estimated by
varying the nominal values by $\pm 1\sigma$. The dilution factor 
fitting range is varied
from nominal $0.00~\cm < |\dz| < 0.20~\cm$ to $0.00~\cm < |\dz| < 0.05~\cm$.
   
For the fitting range for the determination of the final \Asl, the
lower limit is varied from its nominal value of $0.015~\cm$ to $0.040~\cm$ in
several steps.

The results of the systematic error determination for \Asl are
summarized in Table~\ref{sys}.

%\begin{table}[htb]
%  \begin{center}
%    \caption{\label{sys}
%      Source of systematic errors for the measurement of \Asl}
%    \begin{tabular}{lll}
%      \hline\hline
%      category   & source   &  $\Delta \Asl$ (\%)\\
%      \hline
%      event selection & track selection & $\pm0.236$\\
%      & $\cos\theta^*_{\ell\ell}$ cut  & $\pm0.107$\\
%      & lepton pair veto  & $\pm0.167$\\
%      \hline
%      continuum subtraction & & $\pm0.314$\\
%      \hline
%      track corrections & track finding efficiency & $\pm 0.003$\\
%      & electron identification efficiency 
%      & $\pm0.008$\\
%      & muon identification efficiency & $\pm 0.010$\\
%      & fake electrons & $\pm 0.116$\\
%      & fake muons & $\pm 0.489$\\
%      & relative multiplicity &  $\pm 0.104$\\
%      \hline
%      \dz fit for dileptons & detector response function & $\pm 0.009$  \\
%      & \dm & $\pm 0.011$ \\
%      &\taub   & $\pm 0.009$  \\
%      & $80~\micron$ smearing of background \dz & $\pm0.019$ \\
%      & statistics of background MC & $\pm0.022$ \\
%      & fitting range  &  $\pm 0.004$\\
%      &  assuming $N_b^{++} = N_b^{--}$ & $\pm0.139$ \\
%      \hline
%      \dz fit for \Asl & fitting range & $\pm 0.207$ \\
%      \hline
%      total  &   &  $\pm0.722$ \\
%    \end{tabular}
%  \end{center}
%\end{table}
\begin{table}[htb]
  \begin{center}
    \caption{\label{sys}
      Source of systematic errors for the measurement of \Asl}
    \begin{tabular}{lll}
      \hline\hline
      category   & source   &  $\Delta \Asl$ (\%)\\
      \hline
      event selection & track selection & $\pm0.236$\\
      & $\cos\theta^*_{\ell\ell}$ cut  & $\pm0.107$\\
      & lepton pair veto  & $\pm0.167$\\
      \hline
      continuum subtraction & & $\pm0.314$\\
      \hline
      track corrections & track finding efficiency & $\pm 0.074$\\
      & electron identification efficiency 
      & $\pm0.058$\\
      & muon identification efficiency & $\pm 0.208$\\
      & fake electrons & $\pm 0.031$\\
      & fake muons & $\pm 0.047$\\
      & relative multiplicity &  $\pm 0.057$\\
      \hline
      \dz fit for dileptons & detector response function & $\pm 0.009$  \\
      & \dm & $\pm 0.011$ \\
      &\taub   & $\pm 0.009$  \\
      & $68~\micron$ smearing of background \dz & $\pm0.009$ \\
      & statistics of background MC & $\pm0.022$ \\
      & fitting range  &  $\pm 0.004$\\
      &  assuming $N_b^{++} = N_b^{--}$ & $\pm0.139$ \\
      \hline
      \dz fit for \Asl & fitting range & $\pm 0.207$ \\
      \hline
      total  &   &  $\pm0.561$ \\
    \end{tabular}
  \end{center}
\end{table}

\section{Conclusion}
The charge asymmetry for same-sign dilepton events \ups decays has
been measured. The result is related to a $CP$-violation parameter in
\Bz-\BzB mixing, $\Asl=(-0.13 \pm 0.60(\text{stat}) \pm
0.56(\text{sys}))\%$, or equivalently $|q/p| =1.0006 \pm
0.0030(\text{stat}) \pm 0.0028(\text{sys})$. The measured \Asl is
consistent with zero, or equivalently, $|q/p|$ is consistent with
unity. This implies $CP$-violation in \Bz-\BzB mixing is below
the $O(10^{-2})$ level. The $CP$-violation parameter $\epsilon_\B$
can be calculated as $Re(\epsilon_B)/(1+|\epsilon_B|^2) = (-0.3 \pm
1.5(\text{stat}) \pm1.4(\text{sys}))\times 10^{-3}$,
using  the exact formula
\begin{equation}
\frac{Re(\epsilon_B)}{1+|\epsilon_B|^2}
=0.5 \frac{1-\sqrt{(1-A_{sl})/(1+A_{sl})}}
          {1+\sqrt{(1-A_{sl})/(1+A_{sl})}}
%=\frac{1-|q/p|^2}{2(1-|q/p|^2)}
.
\end{equation}
These results are
consistent with and provide significantly more restrictive bounds than previous
measurements~\cite{epsilon}.

\section*{Acknowledgments}
We thank the KEKB group for the excellent operation of the
accelerator, the KEK Cryogenics group for the efficient
operation of the solenoid, and the KEK computer group and
the National Institute of Informatics for valuable computing
and Super-SINET network support. We acknowledge support from
the Ministry of Education, Culture, Sports, Science, and
Technology of Japan and the Japan Society for the Promotion
of Science; the Australian Research Council and the
Australian Department of Education, Science and Training;
the National Science Foundation of China under contract
No.~10175071; the Department of Science and Technology of
India; the BK21 program of the Ministry of Education of
Korea and the CHEP SRC program of the Korea Science and
Engineering Foundation; the Polish State Committee for
Scientific Research under contract No.~2P03B 01324; the
Ministry of Science and Technology of the Russian
Federation; the Ministry of Education, Science and Sport of
the Republic of Slovenia; the National Science Council and
the Ministry of Education of Taiwan; and the U.S.\
Department of Energy.
\\

%\newpage

\end{document}

%% file: author-conf2004.tex
%%% Paper:    
%%% Journal:  summer 2004 conference papers (PRL format)
%%% Contacts: 
%%% Last revised on July 14, 2004 16:40:00 EDT
%%% Non-responding authors or those who said NO are commented out.
%%% ====================================================================
%%% Click the RELOAD button on your web browser to see the updated file.
%%% ====================================================================
%%% Use \input{author} to insert this material into your latex file.
%%%%% Force institutions to appear in alphabetical order when typeset.
\affiliation{Aomori University, Aomori}
\affiliation{Budker Institute of Nuclear Physics, Novosibirsk}
\affiliation{Chiba University, Chiba}
\affiliation{Chonnam National University, Kwangju}
\affiliation{Chuo University, Tokyo}
\affiliation{University of Cincinnati, Cincinnati, Ohio 45221}
\affiliation{University of Frankfurt, Frankfurt}
\affiliation{Gyeongsang National University, Chinju}
\affiliation{University of Hawaii, Honolulu, Hawaii 96822}
\affiliation{High Energy Accelerator Research Organization (KEK), Tsukuba}
\affiliation{Hiroshima Institute of Technology, Hiroshima}
\affiliation{Institute of High Energy Physics, Chinese Academy of Sciences, Beijing}
\affiliation{Institute of High Energy Physics, Vienna}
\affiliation{Institute for Theoretical and Experimental Physics, Moscow}
\affiliation{J. Stefan Institute, Ljubljana}
\affiliation{Kanagawa University, Yokohama}
\affiliation{Korea University, Seoul}
\affiliation{Kyoto University, Kyoto}
\affiliation{Kyungpook National University, Taegu}
\affiliation{Swiss Federal Institute of Technology of Lausanne, EPFL, Lausanne}
\affiliation{University of Ljubljana, Ljubljana}
\affiliation{University of Maribor, Maribor}
\affiliation{University of Melbourne, Victoria}
\affiliation{Nagoya University, Nagoya}
\affiliation{Nara Women's University, Nara}
\affiliation{National Central University, Chung-li}
\affiliation{National Kaohsiung Normal University, Kaohsiung}
\affiliation{National United University, Miao Li}
\affiliation{Department of Physics, National Taiwan University, Taipei}
\affiliation{H. Niewodniczanski Institute of Nuclear Physics, Krakow}
\affiliation{Nihon Dental College, Niigata}
\affiliation{Niigata University, Niigata}
\affiliation{Osaka City University, Osaka}
\affiliation{Osaka University, Osaka}
\affiliation{Panjab University, Chandigarh}
\affiliation{Peking University, Beijing}
\affiliation{Princeton University, Princeton, New Jersey 08545}
\affiliation{RIKEN BNL Research Center, Upton, New York 11973}
\affiliation{Saga University, Saga}
\affiliation{University of Science and Technology of China, Hefei}
\affiliation{Seoul National University, Seoul}
\affiliation{Sungkyunkwan University, Suwon}
\affiliation{University of Sydney, Sydney NSW}
\affiliation{Tata Institute of Fundamental Research, Bombay}
\affiliation{Toho University, Funabashi}
\affiliation{Tohoku Gakuin University, Tagajo}
\affiliation{Tohoku University, Sendai}
\affiliation{Department of Physics, University of Tokyo, Tokyo}
\affiliation{Tokyo Institute of Technology, Tokyo}
\affiliation{Tokyo Metropolitan University, Tokyo}
\affiliation{Tokyo University of Agriculture and Technology, Tokyo}
\affiliation{Toyama National College of Maritime Technology, Toyama}
\affiliation{University of Tsukuba, Tsukuba}
\affiliation{Utkal University, Bhubaneswer}
\affiliation{Virginia Polytechnic Institute and State University, Blacksburg, Virginia 24061}
\affiliation{Yonsei University, Seoul}
  \author{K.~Abe}\affiliation{High Energy Accelerator Research Organization (KEK), Tsukuba} % KEK
  \author{K.~Abe}\affiliation{Tohoku Gakuin University, Tagajo} % TohokuGakuin
  \author{N.~Abe}\affiliation{Tokyo Institute of Technology, Tokyo} % TIT
  \author{I.~Adachi}\affiliation{High Energy Accelerator Research Organization (KEK), Tsukuba} % KEK
  \author{H.~Aihara}\affiliation{Department of Physics, University of Tokyo, Tokyo} % Tokyo
  \author{M.~Akatsu}\affiliation{Nagoya University, Nagoya} % Nagoya
  \author{Y.~Asano}\affiliation{University of Tsukuba, Tsukuba} % Tsukuba
  \author{T.~Aso}\affiliation{Toyama National College of Maritime Technology, Toyama} % Toyama
  \author{V.~Aulchenko}\affiliation{Budker Institute of Nuclear Physics, Novosibirsk} % BINP
  \author{T.~Aushev}\affiliation{Institute for Theoretical and Experimental Physics, Moscow} % ITEP
% ...
% Lausanne
\collaboration{The Belle Collaboration}